\documentclass[3p,times]{elsarticle}

\usepackage{ecrc}

\volume{00}

\firstpage{1}

\journalname{Physics Procedia}

\runauth{S. J. Blundell et al.}

\jnltitlelogo{Physics Procedia}

\usepackage{amssymb}
\usepackage{bm}

\usepackage[figuresright]{rotating}

\begin{document}

\begin{frontmatter}
\dochead{}

\title{A Bayesian approach to magnetic moment determination using $\mu$SR}

\author[OX]{S. J. Blundell\corref{cor}}
\ead{s.blundell@physics.ox.ac.uk}
\author[OX]{A. J. Steele}
\author[OX]{T. Lancaster}
\author[OX]{J. D. Wright}
\author[ISIS]{F. L. Pratt}

\cortext[cor]{Tel. +44-1865-272347}

\address[OX]{Clarendon Laboratory, Department of Physics, Oxford
  University, Parks Road, Oxford OX1 3PU, UK}
\address[ISIS]{ISIS Facility,
Rutherford Appleton Laboratory,
Chilton,
Oxfordshire OX11 0QX,
United Kingdom}

\begin{abstract}
A significant challenge in zero-field $\mu$SR experiments arises from
the uncertainty in the muon site.  It is possible to calculate the
dipole field (and hence precession frequency $\nu$) at any particular
site given the magnetic moment $\mu$ and magnetic structure.  One can
also evaluate $f(\nu)$, the probability distribution function of $\nu$
assuming that the muon site can be anywhere within the unit cell with
equal probability, excluding physically forbidden sites. Since $\nu$
is obtained from experiment, what we would like to know is
$g(\mu\vert\nu)$, the probability density function of $\mu$ {\it
  given} the observed $\nu$.  This can be obtained from our calculated
$f(\nu/\mu)$ using Bayes' theorem.  We describe an approach to this
problem which we have used to extract information about real systems
including a low-moment osmate compound, a family of molecular magnets,
and an iron-arsenide compound.
\end{abstract}

\begin{keyword}
  Muon-spin rotation \sep Dipole fields \sep
  Muon sites

\end{keyword}
\end{frontmatter}

\section{Introduction}
In a $\mu$SR experiment the muon-spin precession frequency $\nu$ can
be used to deduce the local magnetic field $B$ at the muon site.  If
there are a number of muon sites with different local fields $\{ B_j
\}$, then the $\mu$SR signal can contain a number of components with
frequecies $\{ \nu_j \}$.  In complex systems it can be highly
non-trivial to determine the precise location of the muon site or
sites.  However, it is nevertheless useful to attempt to extract an
estimate of the magnitude of the moment of the magnetic species
producing the local field, even in the face of ignorance of the
location of the muon site or sites.  In this paper we describe a
method to attempt this using Bayesian inference.  The paper is
structured as follows: in Sections 2 and 3 we review the theory of
dipolar fields and Bayes' theorem respectively to provide the
necessary background to the calculation which is
described in Section 4.

\section{Dipolar fields}
An implanted muon spin precesses around a local magnetic field, 
$\bm{B}_{\rm local}$, with a frequency $\nu=(\gamma_{\mu}/2\pi) \vert
\bm{B}_{\rm local} \vert$, where $\gamma_{\mu}/2\pi=
135.5~\mathrm{MHz\,T}^{-1}$.  
The magnetic field 
$\bm{B}_{\rm local}$ at the muon site is given by
\begin{equation}
\bm{B}_{\mathrm{local}}=\bm{B}_{0}+\bm{B}_{\mathrm{dipole}}+\bm{B}_{\mathrm{L}}+\bm{B}_{\mathrm{demag}}+\bm{B}_{\mathrm{hyperfine}},\end{equation}
where $\bm{B}_{0}$ represents the applied field (zero in the
experiments considered here), $\bm{B}_{\mathrm{dip}}$ is the dipolar
field from magnetic ions, $\bm{B}_{\mathrm{L}}=\mu_{0}\bm{M}/3$ is the
Lorentz field, $\bm{B}_{\mathrm{demag}}$ is the demagnetizing field
from the sample surface and $\bm{B}_{\mathrm{hyperfine}}$ is the
contact hyperfine field caused by any spin density overlapping with
the muon wavefunction.  In antiferromagnets the Lorentz and
demagnetizing fields vanish (in polycrystalline ferromagnets they
cancel to some extent).  The contact hyperfine field is hard to
estimate but we will neglect it.  The remaining term
is the dipolar field $\bm{B}_{\rm dip}$ and is a function of the
muon-site $\bm{r}_\mu$.  It can be written as
\begin{equation}
B_{\rm dip}^\alpha(\bm{r}_\mu) = \sum_i D_i^{\alpha\beta}(\bm{r}_\mu)\,m_i^\beta, 
\label{eq:localfield}
\end{equation}
a sum over the magnetic ions; the
magnetic moment of the $i$th ion is $\bm{m}_i$.  
In Eq.~(\ref{eq:localfield}), 
$D_i^{\alpha\beta}(\bm{r}_\mu)$ is
the dipolar tensor given by
\begin{equation}
D_i^{\alpha\beta}(\bm{r}_\mu) = {\mu_0\over 4\pi R_i^3} \left(
{3 R_i^\alpha R_i^\beta \over R_i^2} - \delta^{\alpha\beta} \right),
\end{equation}
where $\bm{R}_i\equiv (R_i^x,R_i^y,R_i^z)=\bm{r}_\mu-\bm{r}_i$
and $\delta^{\alpha\beta}$ is the Kronecker delta
($\delta^{\alpha\beta}=1$ if $\alpha=\beta$, else $\delta^{\alpha\beta}=0$).
The behaviour of this tensor is dominated by the arrangement of the
nearest-neighbour magnetic ions and leads to a non-zero local magnetic
field for almost all possible muon sites, even in an
antiferromagnetically ordered system \cite{ptra,musr2008}.
The sum in Eq.~(\ref{eq:localfield}) is taken over the infinite lattice,
but it is well known \cite{mckeehan} that this sum converges in such a
way that it is necessary only to sum over points inside a sphere
centred on $\bm{r}_\mu$ with sufficiently large radius.  
An alternative method of calculation is provided by
the method of Ewald summation 
(for details see \cite{bowden}). 

\section{Bayes' theorem}
We recap some elementary probability theory \cite{sivia,ctp}. 
The {\sl conditional probability} $P({\sf A}\vert {\sf B})$ is the probability that
event ${\sf A}$ occurs {\it given} that event ${\sf B}$ has happened.
The {\sl joint probability} $P({\sf A}\cap {\sf B})$ is the probability that
event ${\sf A}$ and event ${\sf B}$ both occur.
The joint probability  $P({\sf A}\cap {\sf B})$ is equal to the probability
that event {\sf B} occurred multiplied by the probability that {\sf A} occurred,
given that {\sf B} did, i.e.,
\begin{equation}
P({\sf A}\cap {\sf B}) = P({\sf A}\vert {\sf B}) P({\sf B}),
\label{eq:acapb1}
\end{equation}
and, equally well,
\begin{equation}
P({\sf A}\cap {\sf B}) = P({\sf B}\vert {\sf A}) P({\sf A}).
\label{eq:acapb2}
\end{equation}
Now consider the case where there are a number of mutually exclusive
events ${\sf A}_i$ such that
\begin{equation}
\sum_i P({\sf A}_i) = 1.
\end{equation}
Then we can write the probability of some other event {\sf X} as
\begin{equation}
P({\sf X}) = \sum_i P({\sf X}\vert {\sf A}_i) P({\sf A}_i).
\label{bayessum}
\end{equation}
In very general terms, one can say that
given some hypothesis {\sf H} 
there usually exists some computational strategy to evaluate the
probability of a particular outcome {\sf O} assuming that hypothesis
to be correct 
(i.e., there is some method to compute the quantity
$P({\sf O}\vert{\sf H})$).  
However,
what you often want to do is the
reverse of this: you know the outcome because it has actually occurred and you want to
choose an explanation out of the possible hypotheses.  In other words,
given
the outcome you want to
know the probability that the hypothesis is true, and the problem is
that
$P({\sf H}\vert{\sf O})$ is typically much more challenging to evaluate.  
  The needed
transformation of
$P({\sf O}\vert{\sf H})$ into $P({\sf H}\vert{\sf O})$ can be
accomplished using {\sl Bayes' theorem} (named after 
Thomas Bayes (1702--1761), although the modern form is due
to Laplace).   This theorem can be stated as
follows:
\begin{equation}
P({\sf A} \vert {\sf B}) = { P({\sf A}) P({\sf B} \vert {\sf A}) \over P({\sf B}) }.
\end{equation}
Here $P({\sf A})$ is called the {\sl prior probability}, since it is the
probability of ${\sf A}$ occurring without any knowledge as to the outcome
of ${\sf B}$.  The quantity which you derive is
$P({\sf A} \vert {\sf B})$, the {\sl posterior probability}.
The proof of Bayes' theorem is very simple: one simply equates 
Eqs.~(\ref{eq:acapb1}) and (\ref{eq:acapb2}) and rearranges.
For the purpose of this paper, we will write Bayes' theorem using
Eq.~(\ref{bayessum}) as
\begin{equation}
P(\mu \vert \nu ) = { P(\mu) P( \nu \vert \mu )  \over \int P ( \nu
  \vert \mu') P(\mu') \,{\rm d}\mu' }.
\label{eq:bayes}
\end{equation}

\section{Calculation}
Since the positive muon seeks out areas of negative charge density,
constraints can be placed on the likely location of stopped muons.
For example, the muon is unlikely to stop close to the positively
charged ions in a system.  In many oxides, muons have been shown to
stop around $0.1\;\mathrm{nm}$ from an O$^{{2}-}$ ion~\cite{brewer}.
In our calculations, we assume a magnetic moment $\mu$ on the magnetic
species in our material and consider a particular magnetic structure.
Positions in the unit cell are then generated at random and, provided
the relevant constraints are satisfied, the dipole field is calculated
at each of them.  After many such randomly generated positions, one
obtains a distribution of dipole fields.  (This distribution sometimes
have sharp features associated with them which are van Hove
singularities \cite{musr2008,vh}.)  The magnitudes of the resulting
dipole fields are then converted into muon precession frequencies, and
the resulting histogram yields the probability density function (pdf)
$f(\nu/\mu)$, evaluated as a function of precession frequency $\nu$
divided by magnetic moment $\mu$ (since the precession frequency
scales with the magnetic moment).  This function $f(\nu/\mu)$
  allows us to evaluate $P(\nu\vert\mu)$ of Eq.~(\ref{eq:bayes}).
We can write 
\begin{equation}
P(\nu\vert\mu)={1\over \mu}f(\nu/\mu).
\label{eq:bayesexpression}
\end{equation}  
The function
$f(\nu/\mu)$ is normalized so that $\int f(\nu/\mu)\,{\rm
  d}(\nu/\mu)=1$, and hence
the factor of ${1\over \mu}$ is needed in
Eq.~(\ref{eq:bayesexpression}) 
so that
$\int P(\nu\vert\mu)\,{\rm d}\nu=1$.
An example of this approach is shown in Fig.~1(a) for a cubic lattice
of antiferromagnetically coupled spins.  The solid line shows the case
when a simple constraint is applied so that the muon is not permitted
to stop at a site closer than a particular critical distance to the
magnetic moments, resulting in a cut-off of the tail at high frequency.

\begin{figure}[h]
  \centerline{\includegraphics[width=0.40\columnwidth]{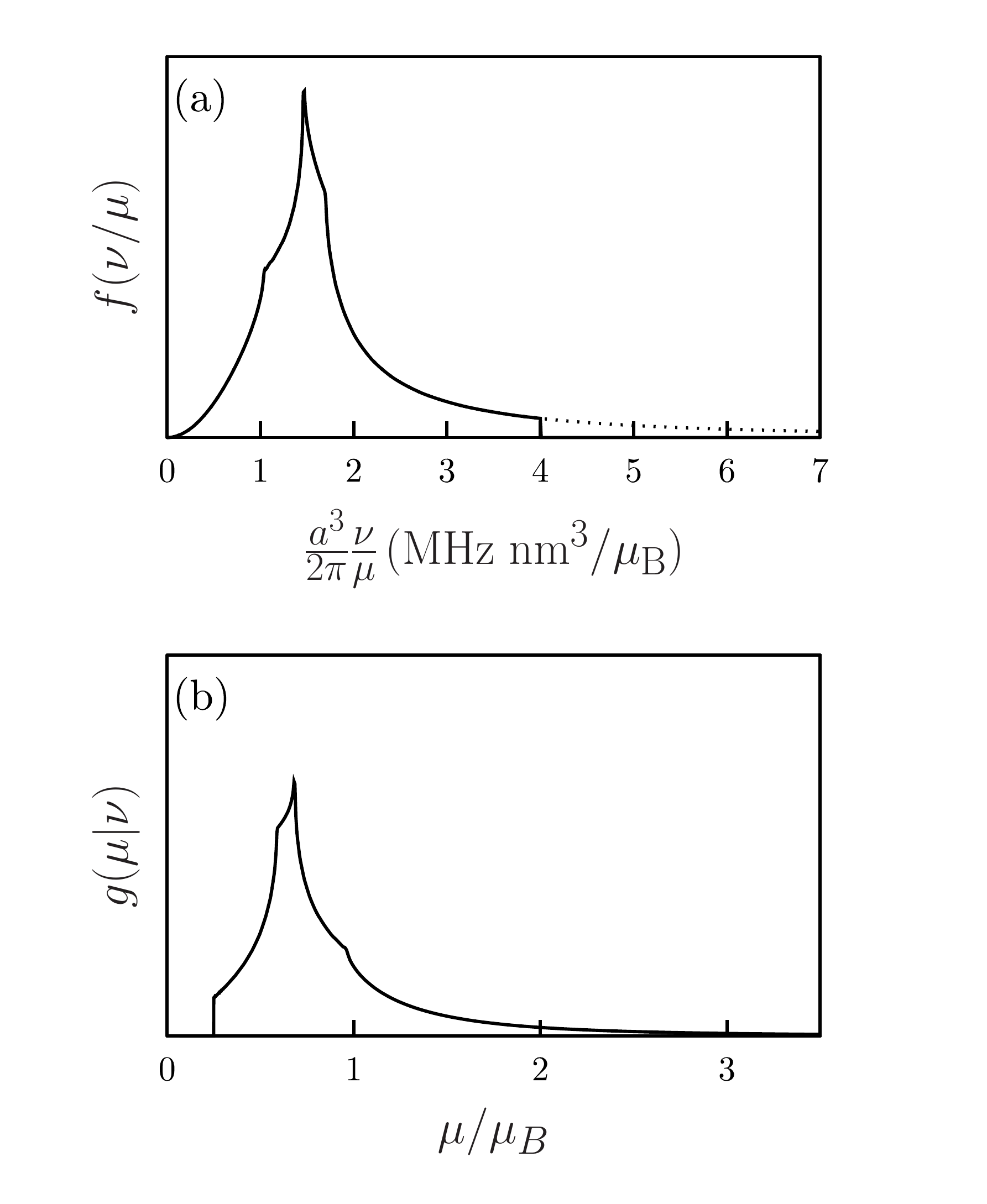}}
  \caption{(a) The dipolar field distribution for a simple cubic
    lattice of antiferromagnetically aligned magnetic moments parallel
    to [110] (with antiferromagnetic wave vector ${\bf
      q}=(\pi,\pi,\pi)$).  The lattice parameter is $a$ and the size
    of the moment is $\mu$ \cite{musr2008}. The solid (dashed) line shows the
    distribution with a cut-off (not) applied.  (b) 
The extracted pdf for the moment given a particular observed
frequency ($\nu=2\pi\mu_{\rm B}/a^3$).    
}
  \label{newplot}
\end{figure}

Since $\nu$ is obtained from a real experiment, what we would like to know is
$g(\mu\vert\nu)$, the pdf of $\mu$ {\it given} the
observed $\nu$.  This can be obtained from our calculated $f(\nu/\mu)$
using Bayes' theorem in the form of Eq.~(\ref{eq:bayes}), which yields
\begin{equation}
g(\mu\vert\nu) = { { 1 \over \mu} f(\nu/\mu) \over \int_0^{\mu_{\rm max}} 
{1 \over \mu'} f(\nu/\mu')\,{\rm d}\mu' }, 
\label{eq:final}
\end{equation}
where we have assumed a prior probability [$P(\mu)$] for the magnetic
moment that is uniform between zero and $\mu_{\rm max}$, and so
$P(\mu)$ is replaced by the uniform probability density $1/\mu_{\rm
  max}$ [which cancels on the top and bottom of Eq.~(\ref{eq:final})].  We
choose $\mu_{\rm max}$ to take a large value, although we have found
that our results are insensitive to the precise value of $\mu_{\rm
  max}$.  A very simple example of this approach is shown in
Fig.~1(b).
When multiple frequencies $\nu_{i}$ are present in the
spectra, it is necessary to multiply their probabilities of
observation in order to obtain the chance of their simultaneous
observation, so we evaluate $g(\mu\vert\{\nu_i\}) \propto \prod_{i}
\int_{\nu_i-\Delta\nu_i}^{\nu_i+\Delta\nu_i}f(\nu_i/\mu)\,{\rm
  d}\nu_i$, where $\Delta\nu_i$ is the error on the fitted frequency.

We have now applied this technique to $\mu$SR data a variety of real systems
in which the muon site is not known.  These include Ba$_2$NaOsO$_6$ in
which we can show from the observed precession frequencies that the
magnetic ground state is most likely to be low-moment ($\approx
0.2$\,$\mu_{\rm B}$) ferromagnetism and not canted antiferromagnetism
\cite{osmate}.  We have also used it to show a reduced moment in the
two-dimensional molecular magnet [Cu(HF$_2$)(pyz)$_2$]BF$_4$ \cite{mm}
and in the pnictide superconductor NaFeAs \cite{jack}.  In all these
cases we do not have a priori information concerning the muon site but
can nevertheless place bounds upon the magnetic moment from the
observed precession signal using this technique.  A possible drawback
that should be borne in mind is that the hyperfine contribution to the
local field is neglected and if this is significant it could affect
the conclusions drawn.  As many of the systems examined so far using
this technique have localized, reduced moments and lower-frequency
precession signals, it is probable that the hyperfine contribution is
not significant in these cases.

\section{Acknowledgments}
We thank EPSRC (UK) for financial support.



\section*{References}
\bibliographystyle{elsarticle-num}

\begin{thebibliography}{00}
\bibitem{ptra} 
S.~J.~Blundell, Phil. Trans. R. Soc. Lond. A357 (1999) 2923.


\bibitem{musr2008}
S. J. Blundell, Physica B 404 (2009) 581.

\bibitem{mckeehan}
L. W. McKeehan, Phys. Rev. 43 (1933) 913. 

\bibitem{bowden}
G. J. Bowden, R. G. Clark, J. Phys. C 14 (1981) L827. 

\bibitem{sivia}
D. S. Sivia, J. Skilling,
  \emph{Data Analysis: A Bayesian Tutorial}
OUP, Oxford (2006), 2nd edn.

\bibitem{ctp}
S. J. Blundell, K. M. Blundell,
  \emph{Concepts in Thermal Physics}
OUP, Oxford (2010), 2nd edn.

\bibitem{brewer}
J. Brewer, R. Kiefl, J. Carolan, P. Dosanjh,W. Hardy,
S. Kreitzman, Q. Li, T. Riseman, P. Schleger,
H. Zhou, et al., Hyp. Int. 63 (1991).

\bibitem{vh}
L. Van Hove, Phys. Rev. 89 (1953) 1189.

\bibitem{osmate}
A. J. Steele, P. J. Baker, T. Lancaster, F. L. Pratt, 
I. Franke, S. Ghannadzadeh, P. A. Goddard, W. Hayes, 
D. Prabhakaran,, S. J. Blundell, Phys. Rev. B 84 (2011) 144416.

\bibitem{mm} 
A. J. Steele, T. Lancaster, S. J. Blundell, P. J. Baker, 
F. L. Pratt, C. Baines, M. M. Conner, H. I. Southerland, 
J. L. Manson, J. A. Schlueter, Phys. Rev. B 84 (2011) 064412.
\bibitem{jack}
J. D. Wright, T. Lancaster, I. Franke, A. J. Steele, J. S. M\"oller, 
M. J. Pitcher, A. J. Corkett, D. R. Parker, S. J. Clarke,
F. L. Pratt, P. J. Baker, S. J. Blundell,
submitted.








\end{thebibliography}

\end{document}